\documentclass[camera ready]{Interspeech}
\title{VISA: A Visual Information Strengthened Audio-Reasoning System for the Interspeech 2026 ARC Agent Track}

\author[affiliation={1,4}, equalcontribution]{Wenming}{Tu}
\author[affiliation={3}, equalcontribution]{Jian}{Gao}
\author[affiliation={1,4}]{Yanru}{Huo}
\author[affiliation={3}]{Yixuan}{Wang}
\author[affiliation={1}]{Jing}{Peng}
\author[affiliation={1}]{Bohan}{Li}
\author[affiliation={1,2}]{Ziyang}{Ma}
\author[affiliation={3}]{Tao}{Liu}
\author[affiliation={3}, correspondingauthor]{Shuai}{Fan}
\author[affiliation={1,3}]{Kai}{Yu}
\author[affiliation={1,2}]{Xie}{Chen}
\author[affiliation={4}, correspondingauthor]{Zilong}{Zheng}

\address{
    $^1$ X-LANCE Lab, Department of Computer Science and Engineering \\
    Shanghai Jiao Tong University, Shanghai, China \\
    $^2$ Shanghai Innovation Institute, Shanghai, China
    $^3$ AISpeech Ltd, Suzhou, China \\
    $^4$ State Key Laboratory of General Artificial Intelligence, BIGAI, Beijing, China 
}

\email{tuwenming@sjtu.edu.cn, shuai.fan@aispeech.com, zlzheng@bigai.ai}

\newcommand{\gray}[1]{\textcolor{gray}{#1}}
\usepackage{tabularx}
\usepackage{booktabs}
\usepackage{makecell}
\usepackage{comment}
\usepackage{multirow}
\usepackage[table]{xcolor}
\usepackage{arydshln}
\usepackage{hyperref}
\usepackage{cleveref}

\begin{document}

\maketitle

\begin{abstract}
Audio reasoning requires multi-step, evidence-grounded inference over temporally dynamic and acoustically mixed signals, exceeding conventional perception tasks such as ASR or captioning. We present VISA, our submission to the Interspeech 2026 Audio Reasoning Challenge (Agent Track), evaluated via the MMAR Rubrics for correctness and reasoning quality. Under a ``LALM as a Tool'' paradigm, VISA strengthe\-ns large audio language models with auxiliary multi-modal evidence while avoiding heavy orchestration. The system integrat\-es three components: multi-modal feature extraction for complementary audio and acoustic-visual clues, model-voting inference with consistency checking for stable predictions, and fine-grained category-aware routing to resolve disagreements and select rubric-aligned reasoning chains. On the official lea\-der\-board, VISA ranks 2nd overall in the Agent Track with a Rubrics score of 66.23\%, while achieving the top Accuracy of 77.40\% in both the Single Model and Agent tracks.
\end{abstract}

\section{Introduction}

Humans routinely infer events, social roles, intentions, and causal relations from speech, music, and ambient sound. Replicating this capability in machines is a core goal of multi-modal AI. Unlike conventional tasks such as ASR~\cite{qwen3asr,peng2026vibevoice}, sound event detection~\cite{shao2024fine,hai2025flexsed}, and audio captioning~\cite{ma2025omni} that mainly evaluate perception, audio reasoning requires multi-step, evidence-grounded inference under mixed sources, temporal dynamics, and noise. While recent Large Audio Language Models (LALMs) excel at auditory understanding, their reasoning remains less reliable and interpretable~\cite{openai2024gpt4ocard,qwen2.5-omni,goel2025audio}.

Recent benchmarks such as MMAR~\cite{ma2025mmar}, MMAU-Pro~\cite{kumar2025mmau}, and Omni-Bench~\cite{li2024omnibench} make this gap explicit. MMAR includes 1,000 real-world audio QA instances spanning sound, music, speech, and their mixtures, and many models still struggle when questions require integrating multiple clues with contextual or domain knowledge. Moreover, benchmarks that emphasize only final accuracy can mask shortcut learning and provide limited visibility into whether the predicted answer is supported by a coherent reasoning process.

\begin{figure*}[h]
    \centering
    \includegraphics[width=\textwidth]{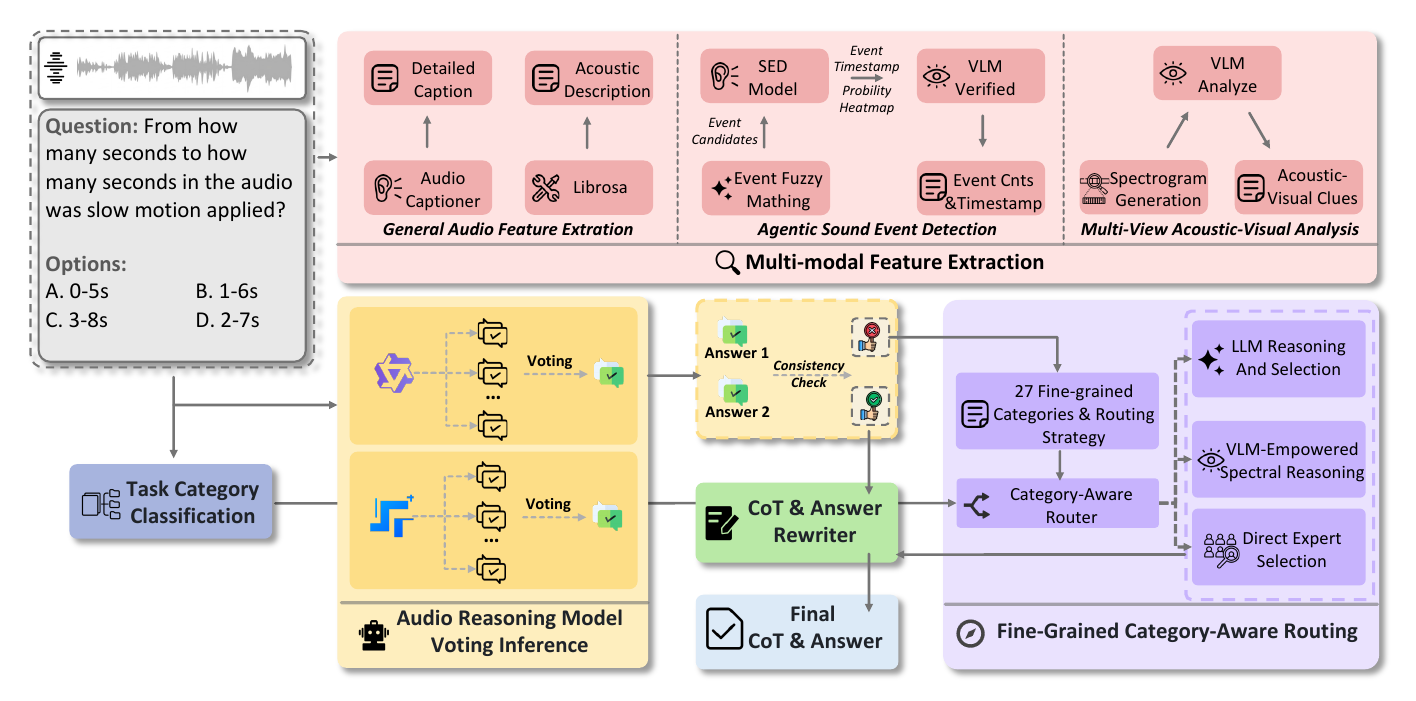}
    \caption{\textbf{Overview of VISA.} The system performs multi-modal feature extraction (audio descriptors, agentic SED, and VLM-based acoustic visual analysis), then conducts multi-model voting with consistency checking, and finally applies fine-grained category-aware routing to select the final CoT and answer.}
    \label{fig:overview}
\end{figure*}

Recent progress in audio reasoning has largely built on the strong perception and understanding of LALMs. Directly answering audio-grounded questions works for short inferences tied to salient clues, but often fails when reasoning requires integrating evidence over time or across mixed sources~\cite{qwen2-audio,salmonn,qwen2.5-omni}. To address this, recent studies enrich supervision with rationale-style signals and apply post-training methods such as scaling and reinforcement learning to encourage more deliberate inference~\cite{goel2025audio,audio-reasoner,audio-thinker}. In parallel, training-free agent systems decompose reasoning into iterative planning, perception, and integration, typically converting audio into intermediate text or structured representations to leverage text LLM reasoning and external tools~\cite{rong2025audiogenie,wijngaard2025audiotoolagent,taheri2025sar}. While agents improve robustness by explicitly collecting and cross-checking evidence, they also introduce orchestration complexity and may produce hallucinated intermediate reasoning. Moreover, transcription-style intermediates can propagate recognition errors or yield ungrounded justifications when incomplete. End-to-end LALMs avoid such tool overhead and operate directly on the original audio, but remain brittle under uncertainty and long-horizon reasoning. Specifically, with the rise of increasingly capable open-source models such as Step-Audio-R1~\cite{tian2025step} and Qwen3-Omni-Thinking~\cite{xu2025qwen3}, it remains unclear how to effectively integrate auxiliary multi-modal evidence into these powerful reasoners to further mitigate their remaining brittleness in complex scenes.

In this report, we present our submission to the \textit{Interspeech 2026 Audio Reasoning Challenge (Agent Track)}\footnote{\url{https://audio-reasoning-challenge.github.io/}}, which evaluates both final answer correctness and the quality of intermediate reasoning chains using the MMAR Rubrics protocol~\cite{ma2026interspeech}. We propose \textbf{VISA}, a system under the ``LALM as a Tool'' paradigm that combines multiple LALMs with auxiliary multi-modal evidence to improve robustness in complex auditory scenes. VISA stabilizes predictions via consensus-based inference and resolves model disagreements with fine-grained, category-aware decision rules. On the official final-stage leaderboard, VISA achieved \textbf{2nd} place overall in the Agent Track with a Rubrics score of \textbf{66.23\%}. Notably, it achieved the highest Accuracy of \textbf{77.40\%} across all submissions in both the Agent and Single Model tracks, demonstrating consistently top-tier correctness together with strong reasoning quality. 

\section{System Design}

\begin{table*}[h]
\centering
\caption{Performance Comparison on MMAR, reported by modality categories and overall average. All numbers are taken from original papers or official benchmarks. The best-performing models in each category are highlighted in \textbf{bold}, and the second-best ones are \underline{underlined}. Results from proprietary models are shown in \textcolor{gray}{gray}.}
\resizebox{\linewidth}{!}{
\begin{tabular}{lccccccccc}
\toprule 
\multirow{2.5}{*}{\textbf{Models}} & \multirow{2.5}{*}{\textbf{Size}} & \multicolumn{3}{c}{\textbf{Single Modality (\%)}} &  \multicolumn{4}{c}{\textbf{Mixed Modalities (\%)}} & \multirow{2.5}{*}{\textbf{Avg (\%)}} \\ 
\cmidrule(lr){3-5} \cmidrule(lr){6-9}
&& \textbf{Sound} & \textbf{Music} & \textbf{Speech} & \textbf{Sound-Music} & \textbf{Sound-Speech} & \textbf{Music-Speech} & \textbf{Sound-Music-Speech} & \\
\midrule
Random Guess & - & 29.39 & 25.88 & 31.48 & 25.00 & 29.30 & 31.10 & 28.13 & 29.32 \\
\midrule \midrule
\multicolumn{10}{c}{\textbf{Large Audio Language Model}} \\ 
\midrule
\gray{GPT-4o Audio~\cite{openai2024gpt4ocard}} & \gray{-} & \gray{53.9} & \gray{51.0} & \gray{70.4} & \gray{63.6} & \gray{72.5} & \gray{62.2} & \gray{\underline{75.0}} & \gray{63.5} \\
\gray{Gemini 2.5 Flash~\cite{google_gemini_25_flash_2025}} & \gray{-} & \gray{60.0} & \gray{53.4} & \gray{77.2} & \gray{63.6} & \gray{76.2} & \gray{\underline{69.5}} & \gray{\underline{75.0}} & \gray{68.4} \\
\gray{Qwen3-Omni Flash Instruct~\cite{xu2025qwen3}} & \gray{-} & \gray{66.7} & \gray{\underline{58.3}} & \gray{74.2} & \gray{\textbf{81.8}} & \gray{78.4} & \gray{64.6} & \gray{70.8} & \gray{69.8} \\
\gray{Qwen3-Omni Flash Thinking~\cite{xu2025qwen3}} & \gray{-} & \gray{65.5} & \gray{51.9} & \gray{75.5} & \gray{63.6} & \gray{76.2} & \gray{65.9} & \gray{70.8} & \gray{68.1} \\
Audio-CoT~\cite{audio-cot} & 7B & 35.8 & 25.2 & 34.0 & 9.1 & 30.7 & 30.5 & 37.5 & 31.3 \\
SALMONN~\cite{salmonn} & 13B & 30.3 & 31.1 & 34.7 & 9.1 & 34.9 & 35.4 & 41.7 & 33.2 \\
Audio-Reasoner~\cite{audio-reasoner} & 7B & 43.6 & 33.5 & 33.0 & 45.5 & 42.7 & 31.7 & 25.0 & 36.8 \\
R1-AQA~\cite{r1-aqa} & 7B & 55.8 & 37.4 & 49.0 & 9.1 & 50.0 & 50.0 & 50.0 & 47.6 \\
Step Audio R1~\cite{tian2025step} & 32B & 61.2 & 52.4 & \underline{81.3} & 54.6 & \underline{80.3} & \textbf{81.7} & \textbf{79.2} & \underline{71.5} \\
Qwen3-Omni Thinking~\cite{xu2025qwen3} & 30B & \underline{67.3} & 51.0 & 77.6 & 54.6 & 78.9 & 73.2 & 70.8 & 69.9 \\
\midrule \midrule
\multicolumn{10}{c}{\textbf{Multi-modal Agent}} \\ 
\midrule
\gray{\textit{AudioToolAgent}~\cite{wijngaard2025audiotoolagent}} & \gray{Agent} & \gray{61.8} & \gray{51.9} & \gray{77.6} & \gray{\underline{72.7}} & \gray{76.6} & \gray{72.0} & \gray{70.8} & \gray{68.8} \\
\textit{AudioToolAgent-Open}~\cite{wijngaard2025audiotoolagent} & Agent & 59.4 & 45.6 & 67.3 & 54.6 & 70.6 & 59.8 & 70.8 & 61.7 \\
AudioGenie-Reasoner~\cite{audio-reasoner} & Agent & 49.7 & 43.3 & 69.2 & 45.5 & 64.5 & 65.3 & 59.1 & 58.9 \\
SAR-LM~\cite{taheri2025sar} & Agent & 52.7 & 56.3 & 81.0 & - & - & - & - & 69.3 \\
\cdashline{1-10}

VISA\textit{(ours)} & Agent & \textbf{71.5} & \textbf{62.6} & \textbf{84.0} & 63.6 & \textbf{86.2} & \textbf{81.7} & \underline{75.0} & \textbf{77.4} \\
\bottomrule
\end{tabular}}
\label{tab:main_results_mmar}
\end{table*}

As illustrated in~\Cref{fig:overview}, VISA consists of three key components. \textbf{(1) Multi-modal Feature Extraction} summarizes audio using acoustic descriptors and auxiliary visual clues (e.g., spectrogram analysis via a VLM) to capture fine-grained details, temporal dynamics, and spatial clues. \textbf{(2) Audio Reasoning Model Voting Inference} queries multiple LALMs and applies consistency-based voting to produce stable, high-confidence predictions. \textbf{(3) Fine-Grained Category-Aware Routing} leverages sub-category heuristics to resolve disagreements by selecting the most suitable output, fully exploiting each model’s strengths and improving overall robustness.
\subsection{Multi-modal feature extraction}
Reliable audio reasoning necessitates evidence that is both comprehensive and easily interpretable by LALMs. To this end, we extract complementary clues from three distinct perspectives: general acoustic descriptions, question-driven event timestamps, and VLM-interpreted acoustic-visual clues. This multi-faceted approach provides a robust and cross-checkable foundation for downstream inference.

\textbf{General Audio Feature Extraction.} To comprehensively represent audio content, we adopt a dual-level extraction strategy. We first utilize the librosa\footnote{\url{https://librosa.org/}} library to extract low-level acoustic descriptions across the temporal, spectral, and cepstral domains, capturing key physical attributes such as energy distribution, timbre (e.g., spectral centroid, bandwidth), and harmonic structures (e.g., MFCCs, HPSS). Complementing these objective signals, we employ the Qwen3-Omni-Captioner~\cite{ma2025omni} to generate high-level semantic descriptions. These captions explicitly articulate auditory events, speaker characteristics, and environmental contexts, thereby translating raw audio signals into interpretable linguistic evidence for subsequent reasoning.

\textbf{Agentic Sound Event Detection.} To extract critical auditory events relevant to the question, we propose an \textbf{Agentic SED} workflow that dynamically identifies and localizes events by integrating query context, audio captions, and answer choices. This process begins with an LLM that extracts potential event candidates from the captions and maps them to standard AudioSet labels via fuzzy matching, ensuring compatibility with the detection model. Subsequently, the FlexSED~\cite{hai2025flexsed} model generates initial detection timestamps and probability heatmaps. Crucially, a Vision Language Model (e.g., Qwen3-VL~\cite{bai2025qwen3}) then visually inspects these heatmaps to verify confidence patterns, correcting fragmented or merged detections. This LLM-guided, VLM-verified pipeline substantially improves both the relevance of detected events and the precision of their temporal localization, which in turn benefits answers to queries that depend on event counts and accurate event timing.

\textbf{Multi-View Acoustic-Visual Analysis.} We observe that LALMs often struggle with precise temporal localization and fine-grained signal perception, particularly for spatial or dynamic acoustic changes. Acoustic visualizations, however, inherently encode rich time-frequency information where such variations manifest as distinct visual patterns. For instance, a sudden deceleration in tempo appears as a downward shift in frequency components, while the approach or retreat of a sound source is clearly reflected in the rising or falling amplitude of the RMS energy curve, consistent with the Doppler effect. Motivated by these visual first principles, we employ a VLM to analyze five types of visualized acoustic representations (e.g., Mel, CQT, RMS). This visual-acoustic grounding enables the model to explicitly capture these signal-level anomalies and temporal dynamics, thereby mitigating hallucinations and enhancing reasoning accuracy for complex auditory scenes.

\subsection{Audio reasoning model voting inference}





To reduce stochastic hallucinations in single-pass generation and exploit complementary reasoning strengths of heterogeneous models, we adopt a hybrid stochastic--deterministic voting strategy. 
We construct an ensemble $\mathcal{M} = \{M_{\text{Qwen}}, M_{\text{Step}}\}$, corresponding to Qwen3-Omni-Thinking and Step-Audio-R1. 
Given an input tuple $x = (A, Q, \mathcal{C})$, where $A$ denotes audio features, $Q$ the query, and $\mathcal{C}$ the candidate set, each model $M_i \in \mathcal{M}$ independently predicts an answer.

For each $M_i$, inference proceeds in three stages:

\begin{enumerate}
\item \textbf{Stochastic Sampling:} We sample $K=3$ outputs with temperature $\tau>0$, forming a candidate set 
$\mathcal{Y}_i = \{y_i^{(1)}, \dots, y_i^{(K)}\}$ to capture output uncertainty.

\item \textbf{Majority Voting:} Let $Count(c, \mathcal{Y}_i)$ denote the frequency of candidate $c$ in $\mathcal{Y}_i$. 
The provisional prediction is obtained via: 
\begin{equation}
\hat{y}_i = \operatorname*{arg\,max}\limits_{c \in \mathcal{C}} ~\text{Count}(c, \mathcal{Y}_i).
\end{equation}


\item \textbf{Deterministic Fallback:} 
With $K=3$, we use the majority-voted answer $\hat{y}_i$ when two or more sampled outputs agree. 
If all three sampled outputs disagree, we regard the stochastic predictions as unstable and re-run the model using greedy decoding ($\tau=0$).

Formally, the final prediction $y_i^*$ for model $M_i$ is defined as:
\begin{equation}
y_i^* =
\begin{cases}
\hat{y}_i, & \text{if a majority exists}, \\
M_i(x; \tau=0), & \text{otherwise}.
\end{cases}
\end{equation}

\end{enumerate}

This strategy filters out random noise while preserving the correct reasoning path when the model is confident, ensuring each agent delivers its most reliable answer before the subsequent routing stage.

\subsection{Fine-grained category-aware routing}
\begin{table*}[htbp]
\centering
\caption{Comparison of accuracy (\%) across 16 sub-categories grouped by hierarchical layers.}
\label{tab:subcategory_full_results}
\small
\resizebox{\textwidth}{!}{%
\begin{tabular}{lc ccc cccccc ccc cccc}
\toprule
\multirow{2.5}{*}[-12pt]{\textbf{Models}} & \multirow{2.5}{*}[-12pt]{\textbf{Avg}}
  & \multicolumn{3}{c}{\textbf{Signal Layer}} 
  & \multicolumn{6}{c}{\textbf{Perception Layer}} 
  & \multicolumn{3}{c}{\textbf{Semantic Layer}} 
  & \multicolumn{4}{c}{\textbf{Cultural Layer}} \\
\cmidrule(lr){3-5} \cmidrule(lr){6-11} \cmidrule(lr){12-14} \cmidrule(lr){15-18}
& 
& \makecell{Acoustic\\Quality\\Analysis} 
& \makecell{Anomaly\\Detection} 
& \makecell{Audio\\Difference\\Analysis} 
& \makecell{Spatial\\Analysis} 
& \makecell{Temporal\\Analysis} 
& \makecell{Correlation\\Analysis} 
& \makecell{Counting\\and Statistics} 
& \makecell{Music\\Theory} 
& \makecell{Env. Perception\\\& Reasoning} 
& \makecell{Content\\Analysis} 
& \makecell{Emotion and\\Intention} 
& \makecell{Speaker\\Analysis} 
& \makecell{Culture of\\Speaker} 
& \makecell{Imagination} 
& \makecell{Aesthetic\\Analysis} 
& \makecell{Prof. Knowledge\\\& Reasoning} \\
\midrule
Step-Audio-R1        & 71.50 & 61.11 & 70.59 & \textbf{87.50} & 73.33 & 60.71 & 70.00 & 39.39 & 61.90 & 77.18 & 79.28 & 83.33 & 77.08 & 82.69 & 60.00 & 62.50 & 66.20 \\
Qwen3-Omni-Thinking  & 69.90 & 77.78 & 76.47 & 50.00 & 66.67 & 46.43 & 78.00 & 59.60 & 44.44 & 77.85 & 76.64 & 66.67 & 75.00 & 76.92 & 60.00 & \textbf{75.00} & 59.15 \\
\midrule \midrule
VISA\textit{(ours)}        & \textbf{77.40} & \textbf{88.89} & \textbf{82.35} & 62.50 & \textbf{86.67} & \textbf{64.29} & \textbf{80.00} & \textbf{59.60} & \textbf{63.49} & \textbf{80.54} & \textbf{82.24} & \textbf{85.00} & \textbf{87.50} & \textbf{86.54} & \textbf{60.00} & 62.50 & 70.42 \\
\quad \textit{w/o fine-grained category} & 73.30 & 83.33 & 70.59 & 62.50 & 66.67 & 53.57 & 72.00 & 51.52 & 57.14 & 77.85 & 81.25 & 85.00 & 77.08 & 73.08 & 60.00 & 75.00 & \textbf{73.24} \\
\bottomrule
\end{tabular}%
}
\end{table*}

\begin{table}[h]
\centering
\caption{Taxonomy of 27 fine-grained audio reasoning categories and their corresponding routing strategies.}
\label{tab:audio_routing}
\scriptsize 
\renewcommand{\arraystretch}{1.2} 

\renewcommand{\tabularxcolumn}[1]{m{#1}} 

\begin{tabularx}{\columnwidth}{@{} >{\centering\arraybackslash}m{0.42\columnwidth} >{\raggedright\arraybackslash}X @{}}
\toprule
\multicolumn{1}{c}{\textbf{Category}} & \multicolumn{1}{c}{\textbf{Description}} \\
\midrule

\rowcolor{gray!15} \multicolumn{2}{c}{\textbf{LLM Reasoning And Selection}} \\
\midrule
Semantic Logic Reasoning & Logical inference from spoken content. \\
Emotion and Intention Inference & Infer speaker emotion or underlying intent. \\
Music Theory Reasoning & Analyze concepts like chords or scales. \\
Style and Genre Classification & Identify specific musical styles or genres. \\
Language/Culture Recognition & Recognize languages and cultural traits. \\
Multi-speaker Reasoning & Understand turn-taking and dialogue logic. \\
Music Structure Understanding & Identify overarching structural elements. \\
Aesthetic and Quality Judgment & Evaluate audio aesthetics or quality. \\
Temporal Order Reasoning & Reason about sound sequence and overlap. \\
Pattern Change Detection & Detect shifts in rhythm, pitch, or structure. \\
Loudness and Intensity Detection & Judge changes in volume or energy. \\
Multi-source Reasoning & Isolate co-occurring sound sources. \\
Noise and Anomaly Detection & Detect abnormal sounds or noise. \\
\midrule

\rowcolor{gray!15} \multicolumn{2}{c}{\textbf{VLM-Empowered Spectral Reasoning}} \\
\midrule
Event Counting and Statistics & Count sound events and provide statistics. \\
Rhythm and Beat Analysis & Analyze tempo, beat patterns, and rhythm. \\
Instrument Identification & Identify specific instruments or families. \\
Distance and Perspective Est. & Estimate sound source distance and depth. \\
Audio Difference Comparison & Compare differences across audio clips. \\
Pitch and Freq. Identification & Identify pitch, frequencies, or intonation. \\
\midrule

\rowcolor{gray!15} \multicolumn{2}{c}{\textbf{Direct Selection (Qwen3-Omni-Thinking)}} \\
\midrule
Environmental Scene Recognition & Recognize acoustic environment. \\
Speech Content Recognition & Understanding spoken sentences. \\
Speaker Identity Analysis & Identify speaker identity, gender, or age. \\
Timbre and Texture Recognition & Distinguish acoustic timbre and textures. \\
Spatial Localization & Locate origin or track movement. \\
\midrule

\rowcolor{gray!15} \multicolumn{2}{c}{\textbf{Direct Selection (Step-Audio-R1)}} \\
\midrule
Sound Source Identification & Identify the sound source. \\
Acoustic Quality Assessment & Assess clarity or distortion. \\
Duration Estimation & Judge duration of specific acoustic events. \\
\bottomrule
\end{tabularx}
\end{table}

While model ensembling commonly adopts majority voting, our analysis shows that heterogeneous audio models exhibit complementary capability boundaries. Specifically, Step-Audio-R1 performs strongly in acoustic quality assessment and fine-grained difference analysis, whereas Qwen3-Omni-Thinking excels in environmental perception and speaker-related tasks. 
To resolve conflicts when predictions diverge, we introduce a \textbf{Disagree-then-Route} mechanism.

We find that coarse-grained routing under broad task labels obscures fine capability distinctions. For example, ``Aesthetic Analysis'' mixes low-level signal quality evaluation (favoring signal-driven models) with high-level stylistic reasoning (requiring stronger logical inference). 
To address this issue, we conduct a heuristic analysis to identify sub-domains where each model consistently dominates, as well as failure cases caused by modality gaps. 
Based on these observations, we refine the taxonomy into \textbf{27 fine-grained categories} and design three routing strategies~(\Cref{tab:audio_routing}):

\begin{enumerate}
    \item \textbf{LLM Reasoning and Selection:} For high-level reasoning tasks (e.g., \textit{Semantic Logic}, \textit{Emotion}), where perception is reliable but inference is complex, an LLM judge evaluates both models' CoTs with supporting evidence and selects the most coherent answer.

    \item \textbf{VLM-Empowered Spectral Reasoning:} For tasks requiring precise quantification or pattern recognition (e.g., \textit{Counting}, \textit{Rhythm}, \textit{Pitch}), which are susceptible to auditory hallucination, we bypass audio models and delegate prediction to a VLM that directly interprets acoustic-visual clues.

    \item \textbf{Direct Expert Selection:} For perception-dominant tasks, we route queries to the empirically superior experts: Qwen3-Omni-Thinking for environmental and speech-related tasks, and Step-Audio-R1 for source and duration analysis.
\end{enumerate}

\begin{table}[h]
\centering
\caption{Leaderboard results of the Audio Reasoning Challenge. ``Rubrics'' denotes the official CoT quality score, and ``Acc'' denotes the objective accuracy.}
\label{tab:main_results}
\small
\begin{tabular}{lcc}
\toprule
\textbf{Model / Team} & \textbf{Rubrics (\%)} & \textbf{Acc (\%)} \\
\midrule
\multicolumn{3}{c}{\textit{Single Model Track}} \\
\midrule
Team A (1st Place)   & 65.29          & 74.00 \\
Team B (2nd Place)   & 62.55          & 71.00          \\
Team C (3rd Place)   & 62.22          & 71.70          \\
Step-Audio-R1        & 58.76          & 71.50          \\
Qwen3-Omni-Thinking  & 58.41          & 69.90          \\
\midrule \midrule
\multicolumn{3}{c}{\textit{Agent Track}} \\
\midrule
Team D (1st Place)   & \textbf{69.83} & \underline{76.90} \\
Team E (3rd Place)   & 66.09          & 75.10             \\
VISA \textit{(ours)} & \underline{66.23} & \textbf{77.40} \\
\cdashline{1-3}
\quad \textit{w/o fine-grained category} & 62.63 & 73.30 \\ 
\bottomrule
\end{tabular}%
\end{table}

\section{Results}

We submitted VISA to the Interspeech 2026 Audio Reasoning Challenge (Agent Track) and conducted a comprehensive evaluation on the MMAR benchmark in terms of both final prediction accuracy and CoT reasoning quality. Results show that VISA achieves state-of-the-art performance across key metrics, validating the effectiveness of its advanced system architecture.

\textbf{Implementation.} For LALMs, we selected Qwen\-3-Omni-Thinking and Step-Audio-R1. We adopt GLM-4.6~\cite{zeng2025glm} as the LLM backbone, due to its strong general-purpose reasoning and instruction-following capabilities. For the analysis of visual information, we utilized Qwen3-VL-235B-A22B~\cite{bai2025qwen3} as our VLM. SED was handled by FlexSED, enabling precise temporal localization and classification of acoustic events. Additionally, we used Qwen3-Omni-Captioner~\cite{ma2025omni} to generate detailed captions describing the acoustic and semantic content of the audio. All components are integrated and configured in accordance with the requirements of the Interspeech 2026 Audio Reasoning Challenge (Agent Track).

\textbf{Comprehensive Performance Analysis.} As shown in~\Cref{tab:main_results_mmar}, VISA achieves an average accuracy of 77.4\% on the MMAR benchmark, outperforming representative agent-based systems such as SAR-LM, the open-source reasoning model Step-Audio-R1, and proprietary models such as GPT-4o-Audio and Gemini 2.5 Flash. This demonstrates that an open-source agent framework can effectively narrow the gap with closed-source models.

\textbf{Effectiveness of Fine-Grained Category-Aware Routing.} As shown in~\Cref{tab:subcategory_full_results} and~\Cref{tab:main_results}, removing fine-grained category-aware routing decreases both overall accuracy and Rubrics score, highlighting the importance of structured routing across 27 sub-categories. Specialized assignment, such as routing anomaly detection to a VLM and duration estimation to dedicated models, improves system robustness. Subcategory results further show gains in Signal Layer, Spatial Analysis, and Temporal Analysis, mainly due to the VLM's ability to extract time-frequency clues from diverse acoustic-visual representations, enabling more precise reasoning over acoustic structure and spatial-temporal dynamics.

\textbf{Challenge Leaderboard Performance and Reasoning Quality.} On the final Agent Track leaderboard~(\Cref{tab:main_results}), VISA ranks 2nd overall with a Rubrics score of 66.23\%, an instance-level metric evaluating the factuality, logical coherence, and completeness of reasoning paths. It also achieves 77.40\% Accuracy, the highest among all systems listed across both the Single Model and Agent tracks. These results show that VISA provides accurate answers with high-quality, rubric-aligned reasoning.

\section{Conclusion}

We present VISA, a system for complex audio reasoning in the Interspeech 2026 Audio Reasoning Challenge (Agent Track). VISA enhances LALMs with visualized acoustic evidence and structured inference strategies, achieving strong accuracy while maintaining high-quality, rubric-aligned reasoning chains. The combination of multi-modal feature extraction, voting-based consistency checking, and category-aware routing proves effective in stabilizing predictions and improving reasoning reliability. Future work will investigate how acoustic-visual clues strengthen audio understanding and multi-step reasoning in LALMs, and explore adaptive, scalable agent-level coordination for complex audio-centric tasks.

\newpage
\section{Acknowledgments}
This work was supported by the Science and Technology Innovation (STI) 2030-Major Project (2022ZD0208700), National Natural Science Foundation of China  (No. U23B2018), Shanghai Municipal Science and Technology Major Project under Grant 2021SHZDZX0102 and Yangtze River Delta Science and Technology Innovation Community Joint Research Project (2024CSJGG1100).

\section{Generative AI Use Disclosure}
During this work, we utilized LLMs to assist in several aspects of the writing and presentation process. The specific applications of LLMs were as follows:

\begin{enumerate}[leftmargin=1.5em]
    \item \textbf{Grammar and Language Refinement:} LLMs were employed to proofread the manuscript for grammatical errors, spelling mistakes, and awkward phrasing. This use was intended to improve the clarity, readability, and overall quality of the written text.

    \item \textbf{Code Correction and Debugging:} For the source code and algorithms presented in this paper, LLMs were used as a tool to help identify and correct syntax errors, debug logical issues, and suggest potential code optimizations.

    \item \textbf{Assistance in Figure Creation:} LLMs provided support in the generation of figures and diagrams. This included generating plotting scripts (e.g., Python's Matplotlib) and offering suggestions for the effective visual representation of data and concepts.
\end{enumerate}

\textbf{The core scientific contributions, including the research concepts, experimental design, data analysis, and the final conclusions, are entirely the work of the authors.} The role of LLMs was strictly limited to that of an assistive tool to enhance the presentation and accuracy of this work.
\bibliographystyle{IEEEtran}
\bibliography{mybib}

\end{document}